\documentclass{emulateapj}
\usepackage{graphicx}
\usepackage{amsfonts}
\usepackage{pifont}

\def\aap{\mbox{A\&A}}
\def\apjl{\mbox{ApJ}}
\def\apjs{\mbox{ApJS}}

\def\ms{\mbox{$M_{\rm *}$}}
\def\mms{\mbox{$m_{\rm *}$}}
\def\mMW{\mbox{$M_{\rm *,MW}$}}

\def\mh{\mbox{$M_{\rm h}$}}

\def\cmf{\mbox{CSMF}}
\def\gsmf{\mbox{GSMF}}
\def\hmf{\mbox{HMF}}

\def\nh{\mbox{$n_h$}}

\def\msun{\mbox{M$_\odot$}}

\def\shmr{\mbox{CHMR}}
\def\ssmr{\mbox{SSMR}}

\def\csmf{\mbox{CSMF}}

\def\pcf{\mbox{2PCF}}

\def\msub{\mbox{$m_{\rm sub}$}}

\def\vmax{\mbox{$v_{\rm max}$}}

\def\lcdm{\mbox{$\Lambda$CDM}}

\def\Nsubmin{\mbox{$\langle N_{\rm sub}(>m_{\rm sub,min}|\mh)\rangle$}}
\def\Nsub{\mbox{$\langle N_{\rm sub}(>\msub|\mh)\rangle$}}

\defcitealias{YMB09}{YMB09}
\defcitealias{Baldry08}{BGD08}
\defcitealias{LW09}{LW09}
\defcitealias{LW11}{LW11}
\defcitealias{Boylan-Kolchin10}{BK+10}
\defcitealias{RAD13}{RAD13}

\def\ltsima{$\; \buildrel < \over \sim \;$}    
\def\lesssim{\lower.5ex\hbox{\ltsima}}           
\def\gtsima{$\; \buildrel > \over \sim \;$}    
\def\grtsim{\lower.5ex\hbox{\gtsima}}           

\slugcomment{Submitted to ApJ}

\shorttitle{Massive satellites in Milky Way-sized hosts}
\shortauthors{Rodriguez-Puebla, Avila-Reese \& Drory}

\begin{document}


\title{The massive satellite population of Milky Way-sized galaxies}


\author{Aldo Rodr\'iguez-Puebla\altaffilmark{1}, Vladimir Avila-Reese\altaffilmark{1} and  Niv Drory\altaffilmark{1,2}}
\affil{\altaffilmark{1}Instituto de Astronom\'ia, 
Universidad Nacional Aut\'onoma de M\'exico,
A. P. 70-264, 04510, M\'exico, D.F., M\'exico.}
\affil{\altaffilmark{2}McDonald Observatory, The University of Texas at Austin, 1 University Station, Austin, TX 78712-0259, USA}

\email{apuebla@astro.unam.mx}

\begin{abstract} 
Several occupational distributions for satellite galaxies more massive than
$\mms\approx 4\times 10^{7}$ \msun\ around Milky-Way (MW)-sized hosts are presented
and used to predict the internal dynamics of these satellites as a function of \mms.
For the analysis, a large galaxy group mock catalog is constructed on 
the basis of (sub)halo-to-stellar mass relations fully constrained with currently available 
observations, namely the galaxy stellar mass function decomposed into centrals and 
satellites, and the two-point correlation functions at different masses. 
We find that $6.6\%$ of MW-sized galaxies host 2 satellites in the mass range of the Small and
Large Magellanic Clouds (SMC and LMC, respectively).
The probabilities of the MW-sized galaxies to have 1 satellite equal or larger than the LMC or
2 satellites equal or larger than the SMC or 3 satellites equal or larger than
Saggitarius (Sgr) are $\approx 0.26, 0.14$, and $0.14$, respectively. The cumulative
satellite mass function of the MW, $N_s(\ge \mms)$, down to the mass of the Fornax dwarf 
is within the $1\sigma$ distribution of all the MW-sized galaxies. We find that 
MW-sized hosts with 3 satellites more massive than Sgr (as the MW) are among 
the most common cases. However, the most and second most massive satellites 
in these systems are smaller than the LMC and SMC by roughly 
0.7 and 0.8 dex, respectively. We conclude that the distribution $N_s(\ge \mms)$ for MW-sized
galaxies is quite broad, the particular case of the MW being of low frequency but not an outlier. 
The halo mass of MW-sized galaxies correlates only weakly with $N_s(\ge \mms)$. Then, 
it is not possible to accurately determine the MW halo mass by means of its $N_s(\ge \mms)$;
from our catalog we constrain a lower limit of $1.38\times 10^{12}$ \msun\ at the $1\sigma$ level. 
Our analysis strongly suggests that the abundance of massive subhalos should agree with 
the abundance of massive satellites in all MW-sized hosts, i.e. there is not a (massive) satellite 
missing problem for the \lcdm\ cosmology. However, we confirm that the maximum circular velocity, \vmax, 
of the subhalos of satellites smaller than $\mms\sim 10^8$ \msun\ is systematically larger 
than the \vmax\ inferred from current observational studies of the MW bright dwarf satellites; at difference
of previous works, this conclusion is based on an analysis of the overall population of MW-sized galaxies. 
Some pieces of evidence suggest that the issue could refer only to satellite dwarfs but not to central 
dwarfs; then, environmental processes associated to dwarfs inside host halos combined with
SN-driven core expansion should be at the basis of the lowering of \vmax. 

\end{abstract}


\section{Introduction}

According to the current paradigm of cosmic structure formation and evolution,
galaxies form inside Cold Dark Matter (CDM) halos, which grow both by diffuse mass accretion 
and by incorporation of smaller halos that become subhalos. Inside the subhalos
(at least inside the more massive ones) galaxies should also have formed prior to their halo's infall, becoming satellite galaxies.
Therefore, the present-day 
population of satellites around central galaxies is the product of the halo/subhalo
assembly and the survival/destruction history of the the galaxies inside the subhalos. The 
N-body simulations within the context of the \lcdm\ cosmological scenario provide us with the 
subhalo conditional mass function (subHCMF) as a function of host halo mass \mh\ 
\citep[see for recent results, e.g.,][]{Springel+2008,Giocoli+2008,Boylan-Kolchin+2010,Gao+2011,Behroozi+2012}. 
Using this function and statistical models constrained by observations, the central/satellite--halo/subhalo 
mass connection can be established \citep[e.g.,][hereafter \citetalias{RAD13}]{Busha+2011,RAD13}. 
In this way, the abundances of the galaxy satellite population as a function of 
\mh\ can be calculated (satellite conditional stellar mass function, \csmf). 
In this paper, our interest is focused on these abundances for systems 
with a central galaxy of Milky Way (MW) stellar mass, \mMW.

With the advent of large galaxy surveys, some observational statistical studies 
of the satellite abundance of central galaxies, in particular those of MW luminosity or 
mass, have beed published. Several statistical distributions have been determined this way, for instance, the fractions of 
MW-sized galaxies with a given number $N_{s} $ of satellites in the mass range of the Magellanic 
Clouds (MC) or with masses equal or larger than the LMC or the SMC 
\citep[][]{James+2011,Tollerud+2011,Liu+11,Busha+2011b,Robotham+2012}.
A natural question is {\it whether the \lcdm\ scenario makes predictions 
in agreement with these statistical results related to scales smaller than previously probed}. 

The works mentioned above conclude that the MW is a rare case with significantly
more massive (MC-sized) satellites than other galaxies of similar luminosity or mass. Other 
studies determine the average luminosity distribution of bright satellites around centrals 
\citep[][]{Lares+2011,Guo+2011,Wang+2012,
Strigari+2012,Jiang+2012,Sales+2013}. The distribution of the MW bright satellites seems to 
lie above the average found for MW-sized galaxies. In spite of all of these studies, it is 
not yet clear whether the satellite luminosity (mass) distribution of the MW is rare in a 
statistically significant sense. It could be that the MW-sized galaxies have a broad 
range of satellite luminosity distributions, the MC-like case being not particularly frequent but 
not an outlier.

The question on how typical is the MW  satellite mass distribution has acquired 
relevance recently. This distribution, being the best studied one, is used to compare 
with subhalo distributions predicted in the context of the \lcdm\ and alternative cosmological 
scenarios in order to test these scenarios at the smallest scales 
\citep[c.f.][]{Boylan-Kolchin+2011b, Boylan-Kolchin+2012,Lowell+2012,Vogelsberger+2012,Zavala+2013}.
However, such a comparison relies (i) on the hope that the MW satellite \csmf\ is not atypical
and (ii) on the assumed halo mass for the MW (the subhalo abundance strongly depends
on \mh, \citealp[e.g.,][]{Gao+2011,Wang+2012b}).
For example, Boylan-Kolchin et al.\ (2011b,2012) have shown that for a few \lcdm\ halos of 
$\sim 10^{12}$ \msun\ resimulated at very high resolution, there is a significant excess of 
subhalos with too high masses or maximum circular velocities ($\vmax> 25$ km/s)
with respect to what is inferred for the MW satellite population (the so-called
``too big to fail" problem).  By means of an analytical model for generating a large sample
of \lcdm\ halos with their corresponding subhalo populations, \citet{Purcell+2012} propose
that the large variation in the latter among different host halos ameliorates the 
``too big to fail" problem: at least $\sim 10\%$ of their MW-sized halos host subhalo 
populations in agreement with the MW dwarf satellite kinematics. \citet{Wang+2012b} 
suggest that the problem is ameliorated if the MW halo mass is simply less massive 
than is commonly thought, $\mh\lesssim 10^{12}$ \msun.

In all of these works, the main caveats are the way the MW satellite population is put into 
the statistical context, and the way the populations of the predicted subhalos and of  
the observed MW satellites are matched. Here, we attempt to overcome these caveats
by using a large mock catalog of MW-sized galaxies, constructed on the basis of
(sub)halo-to-stellar mass relations fully constrained with currently available 
observations, namely the galaxy stellar mass function, \gsmf, decomposed into 
centrals and satellites, and the projected two-point correlation functions, \pcf s, 
measured at different stellar mass bins (for references see Section 2). While
these observations are complete only down to $\approx 2 \times 10^8$ \msun,
the occupational procedure used to construct the catalog allows one to "extrapolate"
observations down to the stellar masses that match the minimum halo/subhalo masses
considered here. In RDA13 \citep[see also][]{Busha+2011}, a preliminary attempt
of studying the massive satellite population of MW-sized galaxies has been 
presented; however, in that paper the results are given as a function of \mh\ 
instead of \mMW, which introduces freedom to choose the right \mh\ to be used for the MW.
 
Our main result from analyzing the mock catalog is that the \lcdm\ scenario
is statistically consistent with observations regarding the abundances
and internal dynamics of satellites in MW-sized galaxies down to satellite stellar 
masses $\mms\sim 10^8$ \msun. At lower masses, down to the limit of our study 
($\mms\sim 10^7$ \msun), the abundances continue being consistent but the 
internal dynamics of observed dwarf satellites suggest that their subhalos have 
\vmax\ values smaller than those of the \lcdm\ subhalos, 
under the assumption that the \vmax\ of the latter remain the same after galaxy 
formation and evolution. Our conclusions
are not affected by uncertainties on the matching of subhalo-satellite
abundances, on the statistical interpretation of the MW nor on the halo MW mass.
Regarding the latter, we instead find the \mh\ distribution
of the MW analogs \citep[see also][]{Busha+2011b}. 

The layout of this paper is as follows. In Section 2 we briefly describe the semi-empirical
occupational approach for linking galaxies to halos and subhalos and how, by using the
results of this approach, we construct a mock catalog of 2 million central galaxies,
each one with its satellite population down to $\mms\sim 10^7$ \msun. From this catalog, we select
a subsample of about 41000 central galaxies with MW-like stellar masses. 
In Section 3, we present different statistical distributions for the massive satellite population of the 
MW-sized galaxies and compare them to some observational studies. 
We investigate the question of how common the MW satellite mass distribution is in 
\S\S 3.1, while in \S\S\ 3.2 we present the halo mass distribution of the MW analogs.
In Section 4 we present \vmax\ vs.\ stellar mass for the mock galaxy (both satellites and centrals)
and compare with observations. Our conclusions and a discussion 
are given in Section 5.
We adopt cosmological parameter values close to WMAP 7:
$\Omega_\Lambda=0.73$, $\Omega_{\rm M}=0.27$, $h=0.70$, $n_s=0.98$,
and $\sigma_8=0.84$.

\section{The method}

In what follows, we briefly review the semi-empirical approach we use
for connecting galaxies to halos and subhalos of different masses. 
For an extensive presentation of this approach, see Section 2 of 
\citetalias{RAD13}. The approach relies on the 
assumption that the central-to-halo and satellite-to-subhalo 
mass relations (\shmr\ and \ssmr, respectively) are monotonic. 
By parametrizing these mass relations, with their intrinsic scatter included,
one can use the predicted \lcdm\ distinct halo and conditional subhalo 
mass functions (HMF and subHCMF) to generate the halo/subhalo 
occupational distributions both for {\it central and satellite} galaxies. Therefore,
this method encapsulates the main ideas behind the abundance 
matching technique, the halo occupation distribution model, and the 
conditional stellar mass function formalism (\citetalias[][see for references therein]{RAD13}; 
see also \citealp{RDA12}). The advantage of the approach is that
all the relevant observed statistical distributions of {\it central and satellite} galaxies
(the \gsmf\ decomposed into centrals and satellites, the \cmf s, and the \pcf s) are 
consistently related to each other and with the predicted halo/subhalo statistical 
distributions (the HMF and subHCMF).

The outputs of this approach are the \shmr\ and \ssmr, including
their intrinsic scatters, and the satellite \csmf s as {\it a function of halo mass \mh}. 
Here we will use the best constrained \shmr\ and \ssmr\ obtained 
in \citetalias{RAD13}. These relations were (over)constrained by making use of all
the available observational information (data set C in \citetalias{RAD13}):
the central and satellite \gsmf s determined by \citet{YMB09} down to 
$2.5\times 10^8$ \msun\ and the projected  \pcf s determined by \citet{Yang2012} in five
stellar mass bins. For the distinct HMF and subHCMF, the \citet{Tinker+2008} 
and \citet[][see also Boylan-Kolchin et al. 2008; Gao et al. 2011]{Boylan-Kolchin+2010}
fits to cosmological simulations were used, respectively.

\subsection{The galaxy group mock catalog}
\label{mock}

Instead of using the analytical \shmr\ and \ssmr\ directly, we apply these  
functions and their scatters to generate a mock galaxy group catalog. With this
catalog we will explore several statistical satellite distributions that can be compared
with some direct observational determinations given as a function of the 
central stellar mass. The catalog is generated as follows: 

\begin{itemize}
\item From a minimum halo mass of $M_{\rm h,min}=10^{10.5}\msun$,
a population of $2\times 10^6$ halos is sampled from the distinct \hmf. 
Each halo is randomly picked from this function by generating a random number
$U$ uniformly distributed between 0 and 1 and finding the value for \mh\
that solves the equation $\nh(\mh)/\nh(M_{\rm h,min})=U$. Here \nh\ is the 
cumulative distinct \hmf.  

\item To each halo a central galaxy with stellar mass \ms\ is assigned randomly
from the probability distribution $P(\ms|\mh)$, i.e.\ the mean \ms--\mh\ relation
and its intrinsic scatter which is assumed to be lognormal distributed with a width of 
0.173 dex (see \citetalias{RAD13}).

\item To each halo defined by its mass \mh\ a subhalo population above $m_{\rm sub,min}=10^{9}$ 
\msun\ is assigned randomly by assuming a Poisson distribution 
\citep{Kravtsov+2004,Boylan-Kolchin+2010}. 
First, the total number of 
subhalos, $\mathcal{N_{\rm sub}}$, above $m_{\rm sub,min}$ is specified by choosing an 
integer from a Poisson distribution with mean \Nsubmin, where this mean number is taken 
from the subHCMF for the given \mh. Then, the mass \msub\ for each subhalo is assigned by solving 
the equation $\Nsub/\Nsubmin=u$, where $u$ again is a random number uniformly distributed between 
0 and 1. Note that this last step should be repeated $\mathcal{N_{\rm sub}}$ times in order to assign 
subhalo masses to each one of the $\mathcal{N_{\rm sub}}$ subhalos. The lower limit in subhalo 
mass is enough to sample satellite galaxies with stellar masses larger than $\mms\approx 10^{7}\msun$, 
see Fig. 7 of \citetalias{RAD13}.

\item To each subhalo we assign a satellite galaxy with stellar mass \mms, taken
from the probability distribution $P(\mms|\msub)$, i.e.\ the mean \mms--\msub\ relation
and an intrinsic scatter, assumed to be lognormal distributed with a width
equal to the central/halo case (the latter assumption seems to be reasonable, 
see \citetalias{RAD13}). 

\end{itemize}

The mock catalog generated in this way reproduces the observational
statistical functions used to constrain the \shmr\ and \ssmr, namely the
\gsmf\ separated into central and satellite galaxies and the \pcf s in several
mass bins. However, the catalog contains much more information, which
can be thought as an ``extension" as well as an extrapolation to lower masses of the 
observations.  In particular, we can find the overall satellite number distributions down to 
$\mms=10^7$ \msun\ around galaxies of a given stellar mass \ms.

\begin{figure}
\vspace*{-250pt}
\includegraphics[height=6.5in,width=6.5in]{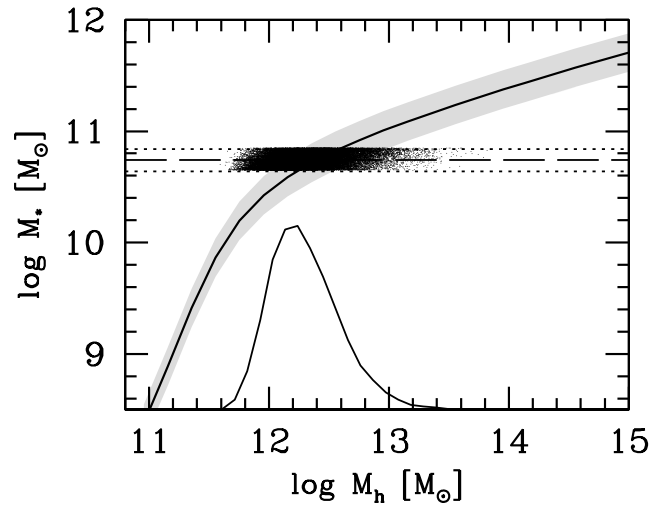}
\caption{Stellar-to-halo mass relation for central galaxies. 
The solid line indicates the \shmr\ reported in \citetalias{RAD13}, while the gray
shaded area shows the $1\sigma$ scatter around the mean, assumed to be 0.173 dex. Galaxies 
that are identified with MW-sized galaxies are those lying in the bin $\log\ms\in[10.64,10.84]$
indicated with the dotted lines. The dashed line indicates the mean of this bin.  The black
dots are the 40694 realizations of MW-sized galaxies. 
The resulting distribution as a function of halo mass for MW-sized galaxies 
is showed below the \shmr. The mean and the standard deviation for this distribution are 
$\log\mh =12.312$ and $0.277$ dex, respectively.
}
\label{MsMh}
\end{figure}

Fig.~\ref{MsMh} illustrates the mean \shmr\ (solid line) and its $1\sigma$ scatter (0.173 dex; gray shaded
area) for the data set C as reported in \citetalias{RAD13}. The 2 million of mock central
galaxies sample this distribution by construction.
The  short dashed line indicates the mass of central galaxies with $\log\ms =10.74$ 
while the dotted lines are 0.1 dex above and below defining a subsample of galaxies 
with stellar masses in the bin $\log$(\mMW/\msun)$\in[10.64,10.84]$, which corresponds to 
the stellar mass estimate for the MW \citep[][]{Flynn+2006}. The 40694 realizations out of the
2 millions that fall within this narrow \ms\ range are represented using black dots 
(\textit{MW-sized galaxies}). The shape of the resulting distribution of this subsample
of central galaxies as a function of \mh\ is shown in the bottom panel of the figure. 
The mean and the standard deviation for this distribution are 
log(\mh/\msun)$=12.312\pm 0.277$. 

\section{Results and comparison to observations}

In the previous Section, we generated a mock catalog of central galaxies corresponding to the
stellar mass estimates for the MW. These galaxies have halos in a 
broad range of masses (see Fig.~1). From this sample, we can then establish the 
fractions (probabilities) of systems with $N_s$ satellites within a (stellar) mass 
range or above a given mass; this mass can not be smaller than $\mms= 10^7$ \msun, 
the minimal mass used to construct the mock catalog (see \S\S \ref{mock}). 
Therefore, our results will be restricted to the population of the largest satellites.

For the statistical calculations, we will assume that the stellar masses
of the LMC and SMC satellite galaxies are $m_{\rm LMC}=2.3\times10^9$ \msun\
and $m_{\rm SMC}=5.3\times 10^8$ \msun\ \citep{James+2011}. We will also
consider that the third most massive MW satellite is Sagittarius (Sgr). For a $V-$band, the
absolute magnitude of $-13.63$ mag and a stellar mass-to-luminosity ratio of 2 for 
Sgr\footnote{We assume for Sgr a stellar population with average metallicity 
[Fe/H]$\approx -0.5$ dex \citep{Chou+2007, Cole+2005} and average age of 8 Gyr
\citep{Bellazzini+2006}.
} imply a stellar mass of
$m_{\rm Sgr} = 5\times 10^7$ \msun. Sgr is a tidally-stripped 
dwarf. Based on observations of its tidal tails, a total magnitude (core + tails) of 
$\approx -15$ mag is obtained \citep{Niederste-Ostholt+2010}. Then, a rough estimate 
of the core + tails stellar mass of Sgr is $m_{\rm Sgr+t}=1.5 \times 10^8$ \msun.
The fourth most massive MW satellite is Fornax (For), with a V-band absolute
magnitude of $-13.3$ mag. An estimate of its stellar mass is $m_{\rm For}= 4.3\times 10^7$ \msun\
 \citep{Boer+2012}

Our mock catalog was constructed based on observational constraints, so the 
different satellite population statistics should be consistent with those of real galaxies;
we expect that this consistency is preserved for the extrapolations to lower masses using this catalog.
In what follows, we compare the results from the mock catalog
with observational distributions of MW-sized galaxies and their population of massive satellites.
It is important to remark that we \textit{do not assume a particular halo mass} for the studied MW-sized
galaxies. 

\begin{figure}
\vspace*{-250pt}
\includegraphics[height=6.5in,width=6.5in]{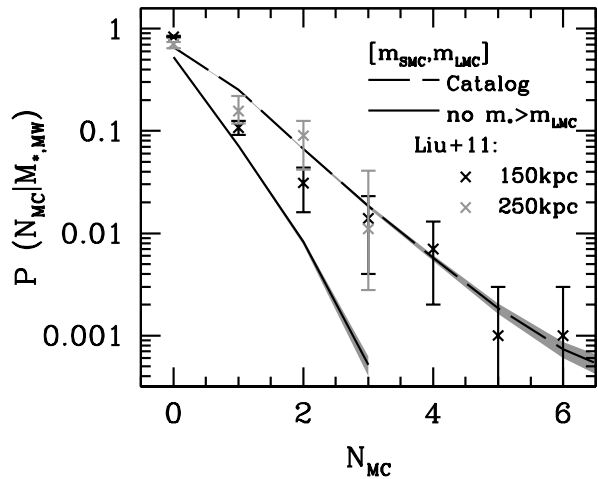}
\caption{Probability of occurrence of $N_{\rm MC}$ satellites in the MC mass range 
around MW-sized galaxies (long-dashed line; solid line is for the extra condition of
no satellites larger than the LMC). The shaded areas are the respective Poissonian errors 
from the counting. Direct observational results from \citep{Liu+11} are plotted 
with black (separation from the host up to 150 kpc) and gray
(separations up to 250 kpc) skeletal symbols.}
\label{pmc}
\end{figure}

From a large SDSS sample, \citet{Liu+11} estimate the fraction of MW-sized isolated 
galaxies that do not have any ($N_{\rm MC} = 0$) and that have $N_{\rm MC}=1,2,3,4,5$ or 6 MC-sized 
satellites. In the same way, we find in our mock catalog the different fractions 
of MW-sized galaxies with $N_{\rm MC}= 0, 1... 6$ satellites in the
stellar mass range $m_{\rm SMC}-m_{\rm LMC}$. 
Figure~\ref{pmc} shows the predicted probabilities (long-dashed line).
The probability of MW-sized galaxies having two MC-sized satellites is $6.6\%$. 
If, from the subsample of MW-like galaxies having $N_{\rm MC}$ MC-sized satellites
we exclude those with satellites larger than the LMC, then the probabilities decrease even 
further (solid line). For $N_{\rm MC}= 2$ and no satellites larger than LMC, the probability 
is now only $0.08\%$. Note that this implies that by far most of those MW-like systems that
have $N_{\rm MC}= 2$ should have at least one satellite more massive than the 
LMC; the MW system does not have such a satellite. 

The results from \citet{Liu+11}, for a search of MC-sized satellites (not excluding
systems with satellites larger than the LMC) up to 150 kpc around the primary, are 
plotted as crosses in Fig.~\ref{pmc}. Note that in our case satellites are 
counted inside the host virial radius ($\sim 200-300$ kpc). Based on Fig.~8 in 
\citet{Liu+11}, we also plot the probabilities when the search radius is 
increased up to 250 kpc (data are provided only for $N_{\rm MC} = 0, 1, 2, 3$). It should 
be said that the selection criteria and observational corrections for searching 
for MC-sized satellites are quite diverse. \citet[][see also \citealp{Busha+2011}]{Liu+11} 
explored the sensitivity of the probabilities to changes in various selection parameters 
and found that their results can slightly change, being the largest sensitivity that one 
to the satellite search radius around the primary. 

\begin{figure}
\vspace*{-250pt}
\includegraphics[height=6.5in,width=6.5in]{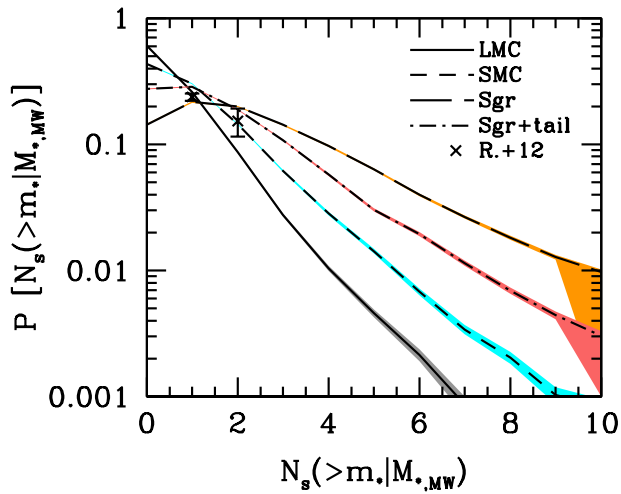}
\caption{Probability of occurrence of $N_s$ satellites around MW-sized galaxies 
with stellar masses equal or larger than the LMC, SMC, Sgr+tail, and
Sgr (solid, short-dashed, dot-short-dashed, and dashed lines, respectively). 
The skeletal symbols are the observational inferences by \citet{Robotham+2012},
corrected for a search radius up to 250 kpc, for satellites equal or more massive than the LMC
and the SMC. }
\label{PNsat}
\end{figure}

The agreement between the probabilities in our mock catalog and the \citet{Liu+11}
observations is good within the uncertainties. It is encouraging that the mock catalog
predicts the statistics of very rare events, as those systems with $N_{\rm MC}\ge 3$, 
in good agreement with observations. Regarding the more common events, in the catalog 
there is a $\sim66\%$ chance of MW-sized
galaxies without MC-sized satellites, while \citet{Liu+11} report 71\% of such galaxies (for
radii up to 250 kpc); this is because we also have slightly more galaxies with $N_{\rm MC} = 1$ than 
in \citet{Liu+11} (the probabilities of systems with more MC-sized satellites are even lower and 
do not contribute significantly).
These small differences can be explained due to the fact that the search radius for satellites in 
\citet{Liu+11} is up to 250 kpc, while in our case there is a fraction of MW-sized galaxies with massive
halos, whose virial radii are larger than 250 kpc. If the satellite search radius would be
increased in \citet{Liu+11}, then the fraction of MW-sized galaxies without MC-sized satellites
would decrease. 

An alternative statistical study of MW analog systems was presented in \citet{Robotham+2012}.
Based on a sample of MW-sized galaxies from the new GAMA survey 
\citep{Driver+2011}, they    
have found the fractions of objects in this sample with one satellite at least as massive as the
LMC or with two satellites at least as massive as the SMC.
From our mock catalog we can calculate the fractions of MW-sized galaxies with any number
of satellites equal or larger than a given stellar mass \mms, $P[N_s(\ge\mms|\mMW)]$. Figure \ref{PNsat}
shows these probabilities for $\mms\ge m_{\rm LMC} $(solid line), $\mms\ge m_{\rm SMC} $
(dashed line), and $\mms\ge m_{\rm Sgr} $ (long-dashed line; the dot-dashed line is for the case when the 
tails of Sgr are included in its mass). The colored contours around the lines are the corresponding 
Poissonian errors from counting. The probabilities of finding one satellite equal or more massive than the LMC
and two satellites equal or more massive than the SMC are $26\%$ and $14.5\%$, respectively. 
In the case of \citet{Robotham+2012} these probabilities are 11.9\% (11.2\%--12.8\%) and  
3.4\% (2.7\% --4.5\%). However, in \citet{Robotham+2012} the satellite search radius 
was fixed to only 70 kpc. From \citet{Liu+11}, we roughly estimate the factors by which these 
fractions could increase if the search radius were to be extended to to 250 kpc; the factors are at least 
2 and 4.5 for $N_{\rm MC} = 1$ and $N_{\rm MC} = 2$, respectively (they could be larger 
because \citet{Liu+11} limit the search to only satellites in the $m_{\rm LMC}-m_{\rm SMC}$ mass range). 
Taking these correction factors into account, the agreement between the predicted probabilities and 
those determined by \citet{Robotham+2012} becomes quite good.

Recently, several authors have measured the complete (bright) satellite abundances around bright centrals, 
in particular those with luminosities close to the MW and M31, by using adequate samples from the SDSS 
\citep[][]{Lares+2011,Guo+2011,Wang+2012,Strigari+2012,Sales+2013} and from the Canada-France-Hawai 
Telescope Legacy Survey \citep{Jiang+2012}. In each one of these studies, different criteria for 
the sample selection, different searching and correction methodologies, various radii for the satellite search, 
etc.\ were applied. Therefore, the results are not easy to compare.

In general,  these works find that the conditional bright satellite luminosity function of MW/M31-sized galaxies 
is described by a relatively steep power law, and a normalization such that down to $\sim 6$ magnitudes 
fainter than the central there is on average a factor of $1.5-2$ fewer satellites than the average of the 
MW and M31. The MW satellite \cmf\  measured in our mock catalog 
agrees in general with the above mentioned studies, but it seems to be slightly overabundant 
above the mass (or luminosity) corresponding to the SMC, in particular with respect to 
\citet{Wang+2012} and \citet{Strigari+2012}. In the case of \citet[][and in a less extent for Guo et al. 2011]{Jiang+2012},
a slight flattening at the high-end of the luminosity function is seen, which is similar to our case. 
We recall that the direct observational searches of satellites are for a fixed radius around the
central, which is 250 or 300 kpc typically (the exception is \citealp{Jiang+2012} who use the virial
radius determined by the \citealp{Yang+2007} group finding algorithm). In the mock catalog we count 
the satellites inside the virial radius, which for a non-negligible fraction of galaxies, is larger than 300 kpc. 
Therefore, it is expected that the number of satellites counted in the direct observational studies (in special of the most
massive ones, which are more probable to be at larger radii) should be slightly lower than in our mock catalog.
 
We conclude that the population of the largest satellites around MW-sized central galaxies in our
mock catalog agrees in general with several direct observational determinations, which present different and limited
satellite population statistics. The advantage of our mock catalog, constrained by observations,
is that allows one to calculate any satellite occupational statistics, and to extend the satellite mass limit 
to masses lower than current direct observational studies. In this way, one may explore in more detail 
how are the satellite populations of MW-sized galaxies and how particular is the MW system.

\subsection{How common is the Milky-Way system?}

According to Fig.~\ref{PNsat}, the MW is less common than similar sized galaxies in the sense
that it has one satellite as massive as the LMC or two satellites equal or more 
massive than the SMC; there are more MW-sized galaxies that do not have satellites of mass 
$\mms\ge m_{\rm LMC}$ (60.6\% vs 26.1\% for those with one satellite) or have less than 
two satellites more massive than $m_{\rm SMC}$ (85.5\% vs 14.5\% for those with two satellites). 
However, the MW can be considered a common galaxy in the sense that
it has three satellites more massive than $m_{\rm Sgr}$. In general, what we learn from 
Fig.~\ref{PNsat} is that the satellite number distributions are relatively wide and there is not
a strongly preferred number of satellites above a given mass. For example, the probabilities
of having 0, 1, 2, 3, or 4 satellites with $\mms\ge m_{\rm Sgr}$ are within a factor of 
less than two from each other.

\begin{figure}
\vspace*{-270pt}
\includegraphics[height=6.5in,width=6.5in]{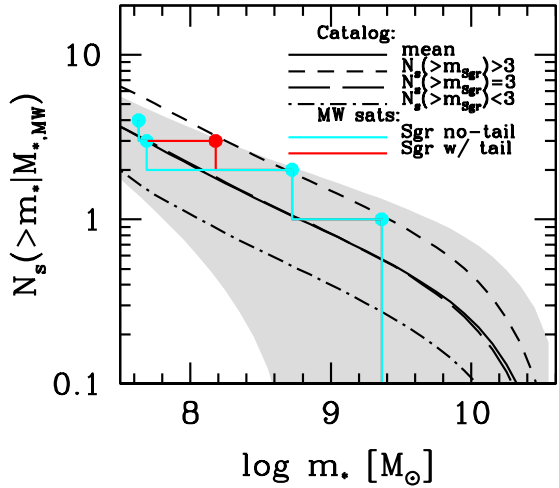}
\caption{Cumulative satellite mass function of MW-sized galaxies in the mock
catalog (solid line) and its 1$\sigma$ scatter (gray shaded area).
Subsamples of  MW-sized galaxies constrained to have $N_s\ge 3$,  $N_s = 3$,
and $N_s\le 3$  satellites more massive than Sgr are indicated with the short-dashed,
long-dashed, and dot-dashed lines, respectively.
The cyan staggered line corresponds to the MW satellite galaxies, while
the red line is for the case the mass of Sgr includes their tidal tails.
}
\label{Nsat}
\end{figure}

The fact that the satellite number distributions of MW-sized galaxies are broad can also be seen
in the plot of the cumulative number of satellites above a given mass \mms\ as a function of \mms, 
$N_s(\ge\mms|\mMW)$, which is related to the satellite \cmf\ discussed 
above. Figure~\ref{Nsat} shows the average (solid line) and the 1$\sigma$ scatter (gray shaded area)
of $N_s(\ge\mms|\mMW)$ from the mock catalog. The latter is quite broad. The cyan 
line corresponds to the MW (the red line is for the case the mass of Sgr includes the tidal tails). 
The MW massive satellite population is within $1\sigma$ of the number distribution
of satellites as a function of mass of all MW-sized galaxies, being above the 
average by less than a factor of 2 at the MC satellite masses, and very close to the average
regarding its three (four) satellites equal or more massive than Sgr (For). By means of 
direct observational determinations
\citet{Guo+2011}, \citet{Strigari+2012}, and \citet{Jiang+2012} arrived 
to a similar conclusion. From a frequency point of view, we find that the MW-sized galaxies 
with one satellite $\ge m_{\rm LMC}$ (two satellites $\ge m_{\rm SMC}$), as the MW, happen 
only 1/0.6=1.68 (2/1.02 =1.92) times less frequently than the average (see also Fig. \ref{PNsat}).

In fact, given that the (massive) satellite number distribution as a function of mass of MW-sized
galaxies is relatively broad, several kind of ``configurations" have close
probabilities and all are relatively low. Besides, as more constraints are imposed
on the configuration (as for example, to have two satellites in the SMC--LMC mass
range but not larger than the LMC, see Fig.~\ref{pmc}), the lower will be the frequency of occurrence. 
However, {\it this does not imply that systems with a particular configuration 
are outliers.} 

In Fig.~\ref{Nsat}, we also show the mean satellite cumulative mass function of the subsamples
of MW-sized galaxies constrained to have $N_s\ge 3$ (short dashed line), $N_s = 3$ (long dashed line),
and $N_s\le 3$ (dot-dashed line) satellites more massive than Sgr. It is interesting to see that \textit{galaxies
with exactly 3 satellites more massive than Sgr are close to the average for MW-sized centrals}, but they 
have typically the most and second most massive satellites smaller than the LMC and SMC by roughly
0.7 and 0.8 dex, respectively. The subsample of galaxies with $N_s\ge 3$ satellites more massive than Sgr 
describes better the satellite mass function of the MW down to the SMC or to Sgr when including its
tails. Finally, we see that the MW definitively does not belong statistically to the subsample
of MW-sized centrals with $N_s\le 3$ satellites more massive than Sgr, contrary to what is 
assumed in \citet{Wang+2012b}.  

The analysis presented above for the MW system can be applied also to M31.
Recent observational results show that M31 has at least twice as many satellites as
does the MW \citep{Yniguez+2013}. 
Specifically, it has six satellites brighter than the luminosity of Sgr, making 
M31 an outlier according to Fig.~\ref{Nsat}. However, the stellar mass of M31 is a factor of $\sim2$ larger
than the MW \citep[e.g.,][]{Tamm+2012}. Therefore, it is expected that the M31 halo is more massive
than the MW one, hence the M31 halo should host more satellites. This question will be analyzed in detail elsewhere.  

\subsection{The halo masses of MW-like systems}

The host halo mass distribution of the MW-sized galaxies in the mock catalog
is plotted in Fig.~\ref{MsMh}. The distribution is broad, with
mean and median values of $2.05\times 10^{12}$ and $1.91 \times 10^{12}$ \msun. 
It is known that for clusters and groups of galaxies the total dynamical mass of the 
system correlates with the richness (number of members above a given
mass, see e.g., \citealp{Reyes+2008}). Is this also the case for MW-like systems? Could we constrain
statistically the MW halo mass by its number of satellites above a given mass
or in between a given mass range?

\begin{figure}
\vspace*{-250pt}
\includegraphics[height=6.5in,width=6.5in]{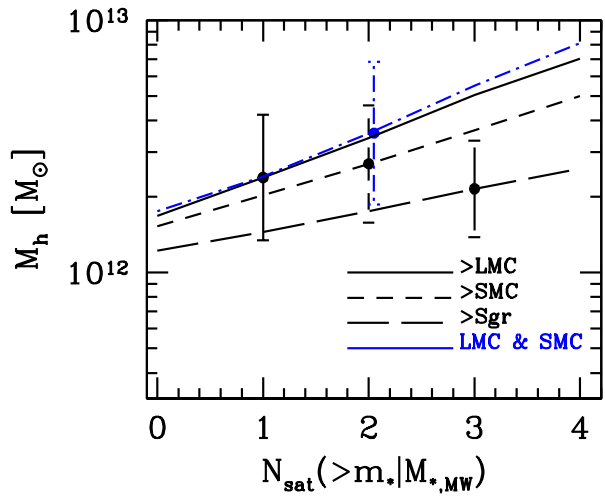}
\caption{Mean host halo mass of the MW-sized galaxies in the mock catalog
as a function of the number of satellites with masses larger or equal than that of the
LMC, SMC, Sgr + tails, and Sgr (see the corresponding lines inside the plot). 
The case for the interval between the LMC and SMC is also included (blue dot-dashed
line). Observe how the halo mass of MW-sized galaxies correlates weakly 
with the number of satellite galaxies. Nevertheless, from all the showed cases,
the MW halo mass is not smaller than $1.38\times 10^{12}$ \msun\ at $1\sigma$ level.
}
\label{MhNsat}
\end{figure}

In Fig.~\ref{MhNsat} we plot the mean host \mh\ of the mock 
MW-systems with $N_s$ satellites with a mass larger or equal than the LMC (solid line), 
SMC (short-dashed line), and Sgr (long-dashed line), and with masses in 
between the SMC and LMC (dot-dashed blue line). The statistical scatter in
all the cases is roughly $\sim0.24$ dex in log\mh. For clarity, we plot
the scatter (vertical lines) only for the cases corresponding to the
MW, i.e., $N_s = 1$ for the solid line, $N_s = 2$ for the short-dashed line, $N_s = 3$ 
for the long-dashed line, and $N_s = 2$ for the dot-dashed blue line (slightly shifted
horizontally).  Figure~\ref{MhNsat} shows that, in general, there is a correlation of \mh~
with $N_s$ but it is weak. The scatter of \mh\  around a given \ms\  does not depend
significantly on $N_s$ for galaxies below the knee in the \ms--\mh\
relation (see Fig.~\ref{MsMh}).

From Fig.~\ref{MhNsat} we can say that {\it at the $1\sigma$ level,
the halo mass of MW-like systems is not smaller than $1.38\times10^{12}$ \msun.} This limit
is for the case of 3 satellites with $\mms\ge m_{\rm Sgr}$ (the mean \mh\ for this case
is log(\mh/\msun)=12.33). Interestingly enough, most of the observational estimates
of the MW halo mass give values above $10^{12}$ \msun. For example, the most recent
work, based on the proper motion of the Leo I dwarf galaxy in combination with numerical
simulations, favors a value of $1.6\times10^{12}\msun$ \citep[][and references therein]{Boylan-Kolchin+2012b}.
For restrictions related to the number of MC-sized 
satellites, the typical halo masses are slightly larger as seen in Fig.~\ref{MhNsat};
for example, log(\mh/\msun) = 12.430$\pm 0.232$ for the case of 2 satellites more
massive than $m_{\rm SMC}$. This estimate is somewhat larger than the one obtained
by \citet{Busha+2011b}, who used the Bolshoi N-body cosmological simulation
\citep{Klypin+2011} for looking for MW-sized halos with two subhalos
with maximum circular velocity, \vmax, larger than 50~km/s (according to our
\vmax--\mms\ relation, this mass corresponds to a smaller  \mms\ than the one 
used here for the SMC, see Fig.~\ref{TF} below; therefore, the host \mh\ estimated 
in \citet{Busha+2011b} would be larger, in better agreement with our study, if they had used 
the \vmax\ corresponding to $m_{\rm Sgr}$). The orbital information of the MC-sized subhalos 
in N-body simulations has been also used for improving the statistical determinations of the 
MW halo mass \citep{Boylan-Kolchin+2011a,Busha+2011b}, finding that the typical masses
should be above log(\mh/\msun) = 12.2-12.3.

\section{Satellite vs \lcdm\ subhalo populations}

The statistical method used to construct our mock catalog allows for a connection 
between satellite and subhalo masses to be made. This connection is constrained by the 
observed satellite \gsmf\ and the projected 
correlation functions at different mass bins (see \citetalias{RAD13}), and it can be 
extrapolated to stellar masses lower than the completeness limit of the observational 
samples. In papers such as Boylan-Kolchin et al.\ (2011b,2012) and \citet{Lowell+2012}, 
the satellite population of the MW is used to discuss the consistency of the 
predicted subhalo population in the \lcdm\ or $\Lambda$WDM scenario, but an
uncertainty remains about whether the MW and their satellites are a typical system
and what the halo mass of the MW is \citep[e.g.,][]{Purcell+2012,Wang+2012b}. 
With our observationally-based catalog, we do not face such a problem since we account
for a large population of MW-sized systems (centrals + satellites), with their corresponding 
host halo masses. 

Our mock catalog offers a statistically complete sample of MW-sized galaxies with
their satellite populations, for which we can "measure" the subhalo masses associated to the 
satellites. By using the tight correlation between maximum circular velocity, \vmax, and mass 
of the subhalos measured in the Millenium-II Simulation \citep[][taking into account the scatter
around this correlation]{Boylan-Kolchin+2010}, the \mms--\vmax\ relation and its scatter 
can be predicted. Note that implicity we are assuming that the subhalo \vmax\  is not altered 
by baryonic effects.
Therefore, in our case, the question is not about a consistency between 
the number of \lcdm\ subhalos (above a given \msub~ or \vmax) and the number of satellites 
(above the \mms\ corresponding to \msub~ or \vmax) --this consistency was stablished by construction 
in the method-- but about whether the predicted \mms--\msub\ or \mms--\vmax\ relations agree
with direct observations.

\begin{figure}
\vspace*{-230pt}
\includegraphics[height=6.1in,width=6.1in]{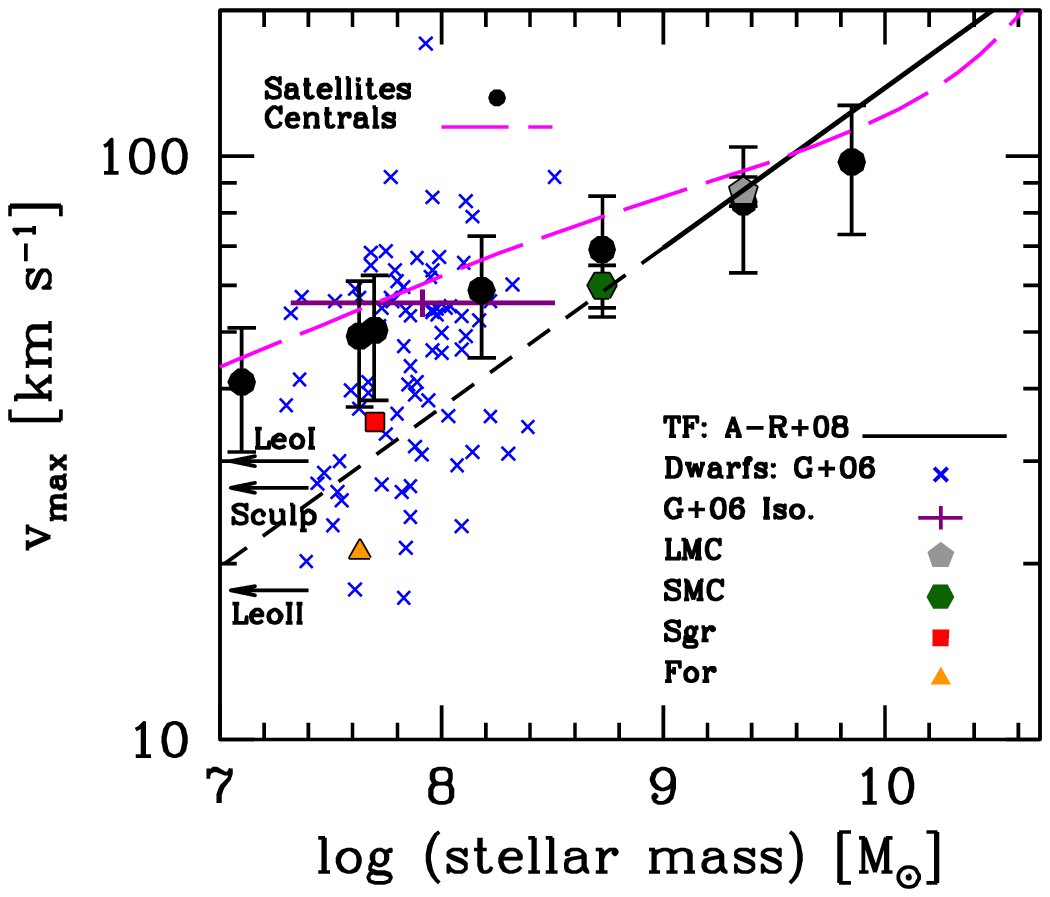}
\caption{The internal dynamics of dwarf galaxies as a function of stellar mass. Solid circles 
with error bars show the mean and the standard deviation of the mock catalog subhalo 
\vmax\ centered at different satellite stellar masses (\mms$=1.2\times10^{7}\msun$,  $m_{*,\rm For}$, 
$m_{*,\rm Sgr}$, $m_{*,\rm SMC}$, $m_{*,\rm LMC}$ and $7.1\times10^{9}\msun$.
The magenta long-dashed line indicates the mean \vmax--\ms\ relation for the mock central galaxies.
Observational estimates for the LMC \citep{Olsen+2011}, SMC \citep{Stanimirovic+2004},
Sgr \citep{Niederste-Ostholt+2010} and For \citep{Strigari+2010} are plotted with the 
color filled symbols. The inferred values of \vmax\ by \citet{Penarrubia+2008} for the next three 
smaller MW dwarfs, Leo I, Sculptor, and Leo II, are indicated with arrows; their stellar masses 
are smaller than $10^7$ \msun.   
The dashed line is an extrapolation to lower masses of the stellar (inverse) TF relation 
of field disk galaxies reported in \citet{Avila-Reese+2008} down to $\sim 10^9$ \msun. 
Individual measurements of $V_{\rm rot}$ and stellar mass for both central and satellite 
dwarfs by \citet[][]{Geha+2006} are plotted with crosses. For a subsample of isolated dwarfs,
\citet{Blanton+2008} report a median $V_{\rm rot}=56\pm 3$ km/s (violet cross). 
}
\label{TF}
\end{figure}

Figure~\ref{TF} shows the mean and standard deviation of the \vmax\ vs.\ \mms\ 
relation for all the satellites above $\mms=10^7$ \msun\ around MW-sized galaxies 
in our mock catalog.  The scatter is due to the dispersions in host halo masses, in 
the \mms--\msub\ relation, and in the \mms--\vmax\ relation.  The dashed line 
is an extrapolation to lower masses of the stellar (inverse) Tully-Fisher (TF) relation 
of field disk galaxies as determined from a suitable catalog in \citet[][the stellar mass was shifted 
by -0.1 dex to convert from the diet-Salpeter to the Chabrier initial mass function]{Avila-Reese+2008}. 
The \vmax--\mms\ relation of the satellites seems to bend towards the low-mass side of the
TF relation of larger galaxies. In fact, a close trend is followed by central galaxies; the solid
line in Fig.~\ref{TF} shows the mean of the  \vmax\ vs.\ \ms\ relation for central galaxies in 
the mock catalog. Such a trend is in agreement with some direct observational studies of the
TF relation of dwarf galaxies \citep[c.f.][]{McGaugh+2000, deRijcke+2007, Amorin+2009}. 
The scatter of the \vmax--\mms\ (as well as the \vmax--\ms) relation increases the lower 
is the mass, also in agreement with direct observational studies. The bend of the stellar TF 
relation at velocities below $\sim 100$ km/s and the increase of its scatter is also observed 
in cosmological numerical simulations \citep{deRossi+2010} and it is explained by the strong 
loss of baryons due to SN-driven feedback in low-amplitude gravitational potentials. 

In Fig.~\ref{TF} we also plot the individual measurements of the maximum rotation velocity $V_{\rm rot}$ 
and stellar mass for (central and satellite) dwarf galaxies by \citet[][crosses]{Geha+2006}\footnote{Note that (i) 
in several cases the HI line widths used to estimate $V_{\rm rot}$ underestimates the real maximum 
velocity that could be at a radius larger than that where gas is observed; and (ii) the galaxy+subhalo \vmax\
after baryon matter is included in the numerical simulations may result significantly lower than in the 
pure dark matter subhalo \vmax\  (see for a discussion Section 5).
}. 
The scatter is high, and down to stellar masses $\sim 10^8$ \msun\ most of dwarfs are close to those
from our catalog and above the extrapolated TF relation. There are some
hints that those dwarfs in the high-$V_{\rm rot}$ side in the \citet[][]{Geha+2006} sample tend 
to be centrals. For example, \citet{Blanton+2008} select the subsample of very isolated dwarfs from 
\citet[][]{Geha+2006}; these are certainly central galaxies. They report a median 
$V_{\rm rot}=56\pm 3$ km/s for this subsample which spans almost all the mass range of the 
total sample. This value is also plotted in Fig.~\ref{TF} (violet cross) and it agrees well with 
the velocities of our central galaxy sample. 

For masses smaller than $10^8$ \msun, there is a significant fraction of observed dwarfs with
lower \vmax\ than the mock dwarfs, although the scatter is high. We also plot the values of 
\vmax\ and \mms\ inferred observationally for the MW satellites LMC, SMC, Sgr, and For  
(color filled symbols; for the sources see the figure caption), as well as the inferred values of 
\vmax\ by \citet{Penarrubia+2008} for the next three smaller MW dwarfs, Leo I, Sculptor, and 
Leo II (indicated with arrows; their stellar masses are certainly smaller than $10^7$ \msun).   
While the LMC and SMC fall close to the mock satellites, 
the observational inferences of \vmax\ for Sgr and For are smaller than the mean
of the mock satellites; a similar difference is expected for the next smaller dwarfs,
Leo I, Sculptor, and Leo II. Even the lower-$1\sigma$ scatter, given mainly by those systems in low-mass
host halos, is higher than the For \vmax.  

For a large sample of galaxies (and not only for the MW galaxy satellites), the results shown above confirm 
{\it a potential problem for the small dwarf galaxies (stellar masses $\lesssim 10^8$ \msun):  
they seem to be associated to significantly less concentrated (smaller \vmax) systems than those 
the \lcdm\ scenario predicts} (Boylan-Kolchin et al. 2011b,2012). However, the 
question that remains open is {\it whether this problem refers to both central and 
satellites galaxies or only to the latter.} According to the above, it could be
that those dwarfs in the \citet[][]{Geha+2006} sample that are in the low-$V_{\rm rot}$ side are
satellites, while those in the high-$V_{\rm rot}$ side are centrals, and as can be appreciated in 
Fig. \ref{TF}, they are consistent with the mock central dwarfs.

\section{Summary \& Discussion}

By means of a statistical approach that observationally constrains the galaxy-(sub)halo 
connection for central and satellite galaxies, we generate a realization of $2\times 10^6$ central 
galaxies and their populations of satellites. Each galaxy is characterized by its stellar and (sub)halo 
mass and, by construction, the catalog reproduces (i) the observed central/satellite \gsmf s and 
projected two-point correlation functions in several stellar mass bins down to their completeness 
limits ($\mms\sim 2.5\times 10^8$ \msun, though we extrapolate it down to $\sim 10^7$ \msun); 
(ii) the \lcdm\ distinct halo mass function \citep{Tinker+2008} as well as the conditional subhalo 
mass functions \citep{Boylan-Kolchin+2010}. From this catalog
we picked all the central galaxies with MW stellar mass, $\log\ms=10.74
\pm 0.1$ dex (40694 objects), and studied the (massive) satellite occupational
distributions of these galaxies. The main results from the ``observation" of the MW-like systems 
in the mock catalog are:

\begin{itemize}

\item The fractions (probabilities) of MW centrals with $N_{\rm MC}$ satellites within the MCs stellar mass 
range or above the SMC or LMC masses are in general agreement with direct observational studies
\citep{Liu+11,Busha+2011,Robotham+2012} after correcting for the satellite search radius, which
in our case is the virial radius of the host halo (see Figs.~\ref{pmc} and \ref{PNsat}).  
For example, we obtain that the probability of finding 2 satellites in the MC mass 
range is $\sim6.6\%$ (or $\sim0.08\%$ if we add the condition of having no satellites
more massive than the LMC); the probabilities of having 1 satellite with $\mms\ge m_{\rm LMC}$,
2 satellites with $\mms\ge m_{\rm SMC}$ or 3 with $\mms\ge m_{\rm Sgr}$ are
$26.1\%$, $14.5\%$, and $14.3\%$, respectively. We also find that the average (massive) satellite mass 
function of the mock MW-sized galaxies is consistent with recent direct observational determinations
of the (bright) conditional satellite luminosity function.

\item Having the two most massive satellites be as
massive as the MCs makes the MW less common, but it is not a rare case in the sense of an outlier. 
In our catalog, MW-sized galaxies with one satellite $\ge  m_{\rm LMC}$ (two satellites $\ge  m_{\rm SMC}$), 
as the MW, happen only 1/0.6=1.68 (2/1.02 =1.92) times less frequently than the average. The cumulative satellite 
mass function of the MW down to the mass of For is within the $1\sigma$ distribution of all the MW-sized galaxies, 
lying above the average by less than a factor of two at the MCs masses and close to the average at the For 
and Sgr masses  (Fig.~\ref{Nsat}). MW-sized centrals with exactly 3 satellites more massive than Sgr are among 
the most common ones, but they have typically the most and second most massive satellites smaller 
than the LMC and SMC by roughly 0.7 and 0.8 dex, respectively. In general, we find that the satellite number 
distributions of MW-sized galaxies are relatively broad.

\item As opposed to clusters and groups of galaxies, the halo mass \mh\ of MW-sized galaxies correlates 
weakly with the number of satellites above a given mass (Fig.~\ref{MhNsat}). The mean log\mh\ and its standard
deviation for galaxies with 3 satellites equal or more massive than Sgr is $12.33\pm0.19$. For 2 satellites
with $\mms\ge  m_{\rm SMC}$ or $\mms$ in the SMC-LMC mass range, the mean and standard
deviation are log\mh$=12.430\pm 0.232$ and log\mh$=12.552\pm 0.283$, respectively. 
Therefore, it is not possible to constrain the halo mass of MW-sized galaxies with appreciable 
accuracy with the satellite population abundance of the MW, but one can say that, at the
$1\sigma$ level, this mass is not smaller than $1.38\times 10^{12}$ \msun, consistent with
recent claims based on the combination of high numerical simulations with the proper motion 
of Leo I \citep{Boylan-Kolchin+2012b}.

\item In our catalog of MW-sized galaxies, the number of \lcdm\ subhalos (above a given \msub\
or \vmax) is consistent with the number of satellites (above the \mms\ corresponding to \msub\
or \vmax) by construction and, being the satellite abundances of the mock galaxies in 
agreement with different direct observational studies, one may conclude that there is not
a (massive) satellite missing problem for the \lcdm\ model. However, we find {\it an internal
dynamics problem}: the \vmax\ of the subhalos of satellites smaller than $\sim 10^8$ \msun\ 
seems to be systematically larger than the \vmax\ 
inferred from current observational studies of dwarf satellites, by factors $\sim1.3-2$ at the
masses of Sgr and For (Fig. \ref{TF}).  There are some hints that this issue could refer only
to the observed satellite dwarfs but not to the central ones. 

\end{itemize}

We conclude that the general agreement of our satellite abundance statistics with direct 
observations signals towards a self-consistency in the construction of the mock catalog, and it shows 
that {\it the underlaying \lcdm\ (sub)halo abundances and internal dynamics are consistent with 
observations down to the scales of the MC galaxies.} For smaller masses, our results point 
out to a possible issue in the internal dynamics of the \lcdm\ (sub)halos as compared with the 
observed satellite dwarfs. These results confirm the conclusions by Boylan-Kolchin et al. (2011b,2012) for
satellite spheroidal dwarfs {\it but with the difference that in our case the results refer to the overall population 
of MW-sized galaxies}. Therefore, our conclusions are free of uncertainties intrinsic to the analysis in
Boylan-Kolchin et al. (2011b,2012) about whether the MW system is atypical 
or not and about what is the MW halo mass to be used \citep[e.g.,][]{Purcell+2012,Wang+2012b}.  
However, before arriving to any conclusion, several aspects of these results should be carefully discussed
(see for an extensive discussion Boylan-Kolchin et al. 2012). Here we highlight two
observational caveats.

(i) Our prediction refers to the maximum circular velocity of the pristine subhalo, \vmax. Observations
refer to the maximum or last-point measured galaxy rotation velocity, $V_{\rm rot}$, or to a model-dependent
\vmax\ constrained by measurements of the stellar velocity dispersion under several assumptions. 
Because dwarf galaxies are dark matter dominated, in the context of the \lcdm\ model it is expected 
that \vmax\ is attained at a radius much larger than the optical one, where the observational tracers 
are not available. Then, it could be that the current observational inferences are underestimating 
the actual values of \vmax. Regarding the dispersion-supported dwarf spheroidals, their unknown 
stellar velocity anisotropy and/or halo shape make ambiguous the inference of their mass distributions 
\citep[e.g.,][and more references therein]{Strigari+2007,Hayashi+2012,Wolf+2012}.

(ii) The mock catalog was constructed using both the \mms--\msub\ and the
\ms--\mh\ relations constrained with
the \citet{YMB09} central and satellite $\gsmf$s down to $\sim 2\times 10^8$, as well as 
with observed projected correlation functions reported in \citet{Yang2012}. For smaller
masses, we use just an extrapolation of this relation and its scatter. If the satellite \gsmf\ at smaller
masses would be steeper than the \citet{YMB09} faint-end or the scatter larger than the assigned by 
us (due, for example, to highly stochastic star formation and tidal effects in the 
satellite dwarfs), then the relation would be shallower and more scattered, which implies lower 
subhalo masses (or \vmax) on average at a given \mms\ and higher scatter in these quantities. 
In \citet{RDA12}, by using a low-mass slope of 
$-1.6$ for the satellite \gsmf, we obtained subhalo masses for $\mms=10^7-10^8$ \msun\ dwarfs 
as small as the tidal masses (close to the subhalo masses) inferred for some MW satellites of 
these masses. This slope is given by \citet{Baldry+2008} for the \gsmf, which goes down to 
$\sim 2.5\times 10^7$ \msun\ after applying a correction for surface brightness incompleteness. 
However, the \gsmf\ in this case refers to all 
galaxies. In \citetalias{RAD13} we decomposed the \citet{Baldry+2008} \gsmf\ into centrals and 
satellites, resulting then in a \mms--\msub\ relation that implies subhalo masses larger than 
the tidal masses by roughly 0.3-0.4 dex. Future samples, complete down to the smallest 
masses and decomposed into central and satellite galaxies, should
tell us whether the satellite \gsmf\ towards very small masses is steep enough or not
as to imply subhalo masses (or \vmax) in better agreement with current dynamical studies. 

Finally, if the observations regarding the faint-end of the satellite \gsmf\ and the internal
dynamics of the dwarf satellites remain roughly as those discussed here, then
our results could be an indication that the baryonic physics significantly affects
the inner structure of the very small subhalos that host dwarf satellites.  A possible 
physical mechanism for explaining the decrease of the inner concentration, and
therefore of \vmax, in low-mass \lcdm\ (sub)halos could be the feedback-driven 
gas outflows. By means of N-body/Hydrodynamics cosmological
simulations, some authors have shown that repeating strong outflows during 
the halo/galaxy growth are able to
drag with them the inner dark matter producing a decreasing of the inner gravitational potential  
\citep[][]{Mashchenko+2008,Maccio+2012,Governato+2012,Zolotov+2012,Ogiya+2012},
though it seems difficult that such an effect would be able to lower \vmax\ to the       
required values \citep{Boylan-Kolchin+2012,diCintio+2011}. However, some numerical simulations
show that {\it in the case of satellites}, the combination of this effect with the stronger 
tidal effects of the halo when a central baryonic galaxy is included, as well
as the lowered baryon fractions of the dwarf satellites,    
work in the direction of reducing the circular velocities of the simulated MW 
satellite dwarf spheroidals to the levels required by the results of  
\citet{Boylan-Kolchin+2012} or our ones \citep[][]{Brooks+2012,Arraki+2012,Gritschneder+2013}.

We have found some hints that the apparent problem of too low-circular velocities
of dwarfs smaller than $\mms\sim 10^8$ \msun\ refers mostly to satellite galaxies 
but not to central ones. If this is the case, then such a problem is explained by the 
plausible physical mechanisms mentioned above. However, if the {\it problem 
remains for isolated dwarfs}, then this could be signaling to the necessity of a modification 
in the cosmological scenario, for example, by introducing warm or self-interacting dark 
matter \citep{Lowell+2012,Vogelsberger+2012}.

\acknowledgments We are grateful to Dr. J. Zavala for thoughtful comments on the draft of this
paper. We also thank to the anonymous referee for a constructive report that helped to improve
this paper.  
A. R-P. acknowledges a graduate student fellowship provided by CONACyT.
V.~A-R.\ and N.~D.\ acknowledge CONACyT (ciencia b\'asica) grants 167332 and 128556.

\bibliographystyle{mn2efix.bst}
\bibliography{blib}

\begin{thebibliography}{68}
\expandafter\ifx\csname natexlab\endcsname\relax\def\natexlab#1{#1}\fi

\bibitem[{{Amor{\'{\i}}n} {et~al}\mbox{.}(2009){Amor{\'{\i}}n}, {Alfonso},
  {Aguerri}, {Mu{\~n}oz-Tu{\~n}{\'o}n}, \& {Cair{\'o}s}}]{Amorin+2009}
{Amor{\'{\i}}n} R., {Alfonso} J., {Aguerri} J.~A.~L., {Mu{\~n}oz-Tu{\~n}{\'o}n}
  C., {Cair{\'o}s} L.~M., 2009, \aap, 501, 75

\bibitem[{{Arraki} {et~al}\mbox{.}(2012){Arraki}, {Klypin}, {More}, \&
  {Trujillo-Gomez}}]{Arraki+2012}
{Arraki} K.~S., {Klypin} A., {More} S., {Trujillo-Gomez} S., 2012, ArXiv
  e-prints

\bibitem[{{Avila-Reese} {et~al}\mbox{.}(2008){Avila-Reese}, {Zavala},
  {Firmani}, \& {Hern{\'a}ndez-Toledo}}]{Avila-Reese+2008}
{Avila-Reese} V., {Zavala} J., {Firmani} C., {Hern{\'a}ndez-Toledo} H.~M.,
  2008, \aj, 136, 1340

\bibitem[{{Baldry}, {Glazebrook} \& {Driver}(2008){Baldry}, {Glazebrook}, \&
  {Driver}}]{Baldry+2008}
{Baldry} I.~K., {Glazebrook} K., {Driver} S.~P., 2008, \mnras, 388, 945

\bibitem[{{Behroozi}, {Wechsler} \& {Conroy}(2012){Behroozi}, {Wechsler}, \&
  {Conroy}}]{Behroozi+2012}
{Behroozi} P.~S., {Wechsler} R.~H., {Conroy} C., 2012, ArXiv e-prints

\bibitem[{{Bellazzini} {et~al}\mbox{.}(2006){Bellazzini}, {Correnti},
  {Ferraro}, {Monaco}, \& {Montegriffo}}]{Bellazzini+2006}
{Bellazzini} M., {Correnti} M., {Ferraro} F.~R., {Monaco} L., {Montegriffo} P.,
  2006, \aap, 446, L1

\bibitem[{{Blanton}, {Geha} \& {West}(2008){Blanton}, {Geha}, \&
  {West}}]{Blanton+2008}
{Blanton} M.~R., {Geha} M., {West} A.~A., 2008, \apj, 682, 861

\bibitem[{{Boylan-Kolchin}, {Besla} \& {Hernquist}(2011){Boylan-Kolchin},
  {Besla}, \& {Hernquist}}]{Boylan-Kolchin+2011a}
{Boylan-Kolchin} M., {Besla} G., {Hernquist} L., 2011, \mnras, 414, 1560

\bibitem[{{Boylan-Kolchin}, {Bullock} \& {Kaplinghat}(2011){Boylan-Kolchin},
  {Bullock}, \& {Kaplinghat}}]{Boylan-Kolchin+2011b}
{Boylan-Kolchin} M., {Bullock} J.~S., {Kaplinghat} M., 2011, \mnras, 415, L40

\bibitem[{{Boylan-Kolchin}, {Bullock} \& {Kaplinghat}(2012){Boylan-Kolchin},
  {Bullock}, \& {Kaplinghat}}]{Boylan-Kolchin+2012}
{Boylan-Kolchin} M., {Bullock} J.~S., {Kaplinghat} M., 2012, \mnras, 422, 1203

\bibitem[{{Boylan-Kolchin} {et~al}\mbox{.}(2012){Boylan-Kolchin}, {Bullock},
  {Sohn}, {Besla}, \& {van der Marel}}]{Boylan-Kolchin+2012b}
{Boylan-Kolchin} M., {Bullock} J.~S., {Sohn} S.~T., {Besla} G., {van der Marel}
  R.~P., 2012, ArXiv e-prints

\bibitem[{{Boylan-Kolchin} {et~al}\mbox{.}(2010){Boylan-Kolchin}, {Springel},
  {White}, \& {Jenkins}}]{Boylan-Kolchin+2010}
{Boylan-Kolchin} M., {Springel} V., {White} S.~D.~M., {Jenkins} A., 2010,
  \mnras, 406, 896

\bibitem[{{Brooks} \& {Zolotov}(2012)}]{Brooks+2012}
{Brooks} A.~M., {Zolotov} A., 2012, ArXiv e-prints

\bibitem[{{Busha} {et~al}\mbox{.}(2011{\natexlab{a}}){Busha}, {Marshall},
  {Wechsler}, {Klypin}, \& {Primack}}]{Busha+2011b}
{Busha} M.~T., {Marshall} P.~J., {Wechsler} R.~H., {Klypin} A., {Primack} J.,
  2011{\natexlab{a}}, \apj, 743, 40

\bibitem[{{Busha} {et~al}\mbox{.}(2011{\natexlab{b}}){Busha}, {Wechsler},
  {Behroozi}, {Gerke}, {Klypin}, \& {Primack}}]{Busha+2011}
{Busha} M.~T., {Wechsler} R.~H., {Behroozi} P.~S., {Gerke} B.~F., {Klypin}
  A.~A., {Primack} J.~R., 2011{\natexlab{b}}, \apj, 743, 117

\bibitem[{{Chou} {et~al}\mbox{.}(2007){Chou}, {Majewski}, {Cunha}, {Smith},
  {Patterson}, {Mart{\'{\i}}nez-Delgado}, {Law}, {Crane}, {Mu{\~n}oz}, {Garcia
  L{\'o}pez}, {Geisler}, \& {Skrutskie}}]{Chou+2007}
{Chou} M.-Y. {et~al.}, 2007, \apj, 670, 346

\bibitem[{{Cole} {et~al}\mbox{.}(2005){Cole}, {Tolstoy}, {Gallagher}, \&
  {Smecker-Hane}}]{Cole+2005}
{Cole} A.~A., {Tolstoy} E., {Gallagher}, III J.~S., {Smecker-Hane} T.~A., 2005,
  \aj, 129, 1465

\bibitem[{{de Boer} {et~al}\mbox{.}(2012){de Boer}, {Tolstoy}, {Hill}, {Saha},
  {Olszewski}, {Mateo}, {Starkenburg}, {Battaglia}, \& {Walker}}]{Boer+2012}
{de Boer} T.~J.~L. {et~al.}, 2012, \aap, 544, A73

\bibitem[{{De Rijcke} {et~al}\mbox{.}(2007){De Rijcke}, {Zeilinger}, {Hau},
  {Prugniel}, \& {Dejonghe}}]{deRijcke+2007}
{De Rijcke} S., {Zeilinger} W.~W., {Hau} G.~K.~T., {Prugniel} P., {Dejonghe}
  H., 2007, \apj, 659, 1172

\bibitem[{{de Rossi}, {Tissera} \& {Pedrosa}(2010){de Rossi}, {Tissera}, \&
  {Pedrosa}}]{deRossi+2010}
{de Rossi} M.~E., {Tissera} P.~B., {Pedrosa} S.~E., 2010, \aap, 519, A89

\bibitem[{{di Cintio} {et~al}\mbox{.}(2011){di Cintio}, {Knebe}, {Libeskind},
  {Yepes}, {Gottl{\"o}ber}, \& {Hoffman}}]{diCintio+2011}
{di Cintio} A., {Knebe} A., {Libeskind} N.~I., {Yepes} G., {Gottl{\"o}ber} S.,
  {Hoffman} Y., 2011, \mnras, 417, L74

\bibitem[{{Driver} {et~al}\mbox{.}(2011){Driver}, {Hill}, {Kelvin}, {Robotham},
  {Liske}, {Norberg}, {Baldry}, {Bamford}, {Hopkins}, {Loveday}, {Peacock},
  {Andrae}, {Bland-Hawthorn}, {Brough}, {Brown}, {Cameron}, {Ching}, {Colless},
  {Conselice}, {Croom}, {Cross}, {de Propris}, {Dye}, {Drinkwater}, {Ellis},
  {Graham}, {Grootes}, {Gunawardhana}, {Jones}, {van Kampen}, {Maraston},
  {Nichol}, {Parkinson}, {Phillipps}, {Pimbblet}, {Popescu}, {Prescott},
  {Roseboom}, {Sadler}, {Sansom}, {Sharp}, {Smith}, {Taylor}, {Thomas},
  {Tuffs}, {Wijesinghe}, {Dunne}, {Frenk}, {Jarvis}, {Madore}, {Meyer},
  {Seibert}, {Staveley-Smith}, {Sutherland}, \& {Warren}}]{Driver+2011}
{Driver} S.~P. {et~al.}, 2011, \mnras, 413, 971

\bibitem[{{Flynn} {et~al}\mbox{.}(2006){Flynn}, {Holmberg}, {Portinari},
  {Fuchs}, \& {Jahrei{\ss}}}]{Flynn+2006}
{Flynn} C., {Holmberg} J., {Portinari} L., {Fuchs} B., {Jahrei{\ss}} H., 2006,
  \mnras, 372, 1149

\bibitem[{{Gao} {et~al}\mbox{.}(2011){Gao}, {Frenk}, {Boylan-Kolchin},
  {Jenkins}, {Springel}, \& {White}}]{Gao+2011}
{Gao} L., {Frenk} C.~S., {Boylan-Kolchin} M., {Jenkins} A., {Springel} V.,
  {White} S.~D.~M., 2011, \mnras, 410, 2309

\bibitem[{{Geha} {et~al}\mbox{.}(2006){Geha}, {Blanton}, {Masjedi}, \&
  {West}}]{Geha+2006}
{Geha} M., {Blanton} M.~R., {Masjedi} M., {West} A.~A., 2006, \apj, 653, 240

\bibitem[{{Giocoli}, {Tormen} \& {van den Bosch}(2008){Giocoli}, {Tormen}, \&
  {van den Bosch}}]{Giocoli+2008}
{Giocoli} C., {Tormen} G., {van den Bosch} F.~C., 2008, \mnras, 386, 2135

\bibitem[{{Governato} {et~al}\mbox{.}(2012){Governato}, {Zolotov}, {Pontzen},
  {Christensen}, {Oh}, {Brooks}, {Quinn}, {Shen}, \&
  {Wadsley}}]{Governato+2012}
{Governato} F. {et~al.}, 2012, \mnras, 422, 1231

\bibitem[{{Gritschneder} \& {Lin}(2013)}]{Gritschneder+2013}
{Gritschneder} M., {Lin} D.~N.~C., 2013, \apj, 765, 38

\bibitem[{{Guo} {et~al}\mbox{.}(2011){Guo}, {Cole}, {Eke}, \&
  {Frenk}}]{Guo+2011}
{Guo} Q., {Cole} S., {Eke} V., {Frenk} C., 2011, \mnras, 417, 370

\bibitem[{Hayashi \& Chiba(2012)}]{Hayashi+2012}
Hayashi K., Chiba M., 2012, The Astrophysical Journal, 755, 145

\bibitem[{{James} \& {Ivory}(2011)}]{James+2011}
{James} P.~A., {Ivory} C.~F., 2011, \mnras, 411, 495

\bibitem[{Jiang, Jing \& Li(2012)Jiang, Jing, \& Li}]{Jiang+2012}
Jiang C.~Y., Jing Y.~P., Li C., 2012, \apj, 760, 16

\bibitem[{{Klypin}, {Trujillo-Gomez} \& {Primack}(2011){Klypin},
  {Trujillo-Gomez}, \& {Primack}}]{Klypin+2011}
{Klypin} A.~A., {Trujillo-Gomez} S., {Primack} J., 2011, \apj, 740, 102

\bibitem[{{Kravtsov} {et~al}\mbox{.}(2004){Kravtsov}, {Berlind}, {Wechsler},
  {Klypin}, {Gottl{\"o}ber}, {Allgood}, \& {Primack}}]{Kravtsov+2004}
{Kravtsov} A.~V., {Berlind} A.~A., {Wechsler} R.~H., {Klypin} A.~A.,
  {Gottl{\"o}ber} S., {Allgood} B., {Primack} J.~R., 2004, \apj, 609, 35

\bibitem[{{Lares}, {Lambas} \& {Dom{\'{\i}}nguez}(2011){Lares}, {Lambas}, \&
  {Dom{\'{\i}}nguez}}]{Lares+2011}
{Lares} M., {Lambas} D.~G., {Dom{\'{\i}}nguez} M.~J., 2011, \aj, 142, 13

\bibitem[{{Liu} {et~al}\mbox{.}(2011){Liu}, {Gerke}, {Wechsler}, {Behroozi}, \&
  {Busha}}]{Liu+11}
{Liu} L., {Gerke} B.~F., {Wechsler} R.~H., {Behroozi} P.~S., {Busha} M.~T.,
  2011, \apj, 733, 62

\bibitem[{{Lovell} {et~al}\mbox{.}(2012){Lovell}, {Eke}, {Frenk}, {Gao},
  {Jenkins}, {Theuns}, {Wang}, {White}, {Boyarsky}, \&
  {Ruchayskiy}}]{Lowell+2012}
{Lovell} M.~R. {et~al.}, 2012, \mnras, 420, 2318

\bibitem[{{Macci{\`o}} {et~al}\mbox{.}(2012){Macci{\`o}}, {Stinson}, {Brook},
  {Wadsley}, {Couchman}, {Shen}, {Gibson}, \& {Quinn}}]{Maccio+2012}
{Macci{\`o}} A.~V., {Stinson} G., {Brook} C.~B., {Wadsley} J., {Couchman}
  H.~M.~P., {Shen} S., {Gibson} B.~K., {Quinn} T., 2012, \apjl, 744, L9

\bibitem[{{Mashchenko}, {Wadsley} \& {Couchman}(2008){Mashchenko}, {Wadsley},
  \& {Couchman}}]{Mashchenko+2008}
{Mashchenko} S., {Wadsley} J., {Couchman} H.~M.~P., 2008, Science, 319, 174

\bibitem[{{McGaugh} {et~al}\mbox{.}(2000){McGaugh}, {Schombert}, {Bothun}, \&
  {de Blok}}]{McGaugh+2000}
{McGaugh} S.~S., {Schombert} J.~M., {Bothun} G.~D., {de Blok} W.~J.~G., 2000,
  \apjl, 533, L99

\bibitem[{{Niederste-Ostholt} {et~al}\mbox{.}(2010){Niederste-Ostholt},
  {Belokurov}, {Evans}, \& {Pe{\~n}arrubia}}]{Niederste-Ostholt+2010}
{Niederste-Ostholt} M., {Belokurov} V., {Evans} N.~W., {Pe{\~n}arrubia} J.,
  2010, \apj, 712, 516

\bibitem[{{Ogiya} \& {Mori}(2012)}]{Ogiya+2012}
{Ogiya} G., {Mori} M., 2012, ArXiv e-prints

\bibitem[{{Olsen} {et~al}\mbox{.}(2011){Olsen}, {Zaritsky}, {Blum}, {Boyer}, \&
  {Gordon}}]{Olsen+2011}
{Olsen} K.~A.~G., {Zaritsky} D., {Blum} R.~D., {Boyer} M.~L., {Gordon} K.~D.,
  2011, \apj, 737, 29

\bibitem[{{Pe{\~n}arrubia}, {McConnachie} \& {Navarro}(2008){Pe{\~n}arrubia},
  {McConnachie}, \& {Navarro}}]{Penarrubia+2008}
{Pe{\~n}arrubia} J., {McConnachie} A.~W., {Navarro} J.~F., 2008, \apj, 672, 904

\bibitem[{{Purcell} \& {Zentner}(2012)}]{Purcell+2012}
{Purcell} C.~W., {Zentner} A.~R., 2012, \jcap, 12, 7

\bibitem[{{Reyes} {et~al}\mbox{.}(2008){Reyes}, {Mandelbaum}, {Hirata},
  {Bahcall}, \& {Seljak}}]{Reyes+2008}
{Reyes} R., {Mandelbaum} R., {Hirata} C., {Bahcall} N., {Seljak} U., 2008,
  \mnras, 390, 1157

\bibitem[{{Robotham} {et~al}\mbox{.}(2012){Robotham}, {Baldry},
  {Bland-Hawthorn}, {Driver}, {Loveday}, {Norberg}, {Bauer}, {Bekki}, {Brough},
  {Brown}, {Graham}, {Hopkins}, {Phillipps}, {Power}, {Sansom}, \&
  {Staveley-Smith}}]{Robotham+2012}
{Robotham} A.~S.~G. {et~al.}, 2012, \mnras, 424, 1448

\bibitem[{{Rodr{\'{\i}}guez-Puebla}, {Avila-Reese} \&
  {Drory}(2013){Rodr{\'{\i}}guez-Puebla}, {Avila-Reese}, \& {Drory}}]{RAD13}
{Rodr{\'{\i}}guez-Puebla} A., {Avila-Reese} V., {Drory} N., 2013, \apj, 767, 92

\bibitem[{{Rodr{\'{\i}}guez-Puebla}, {Drory} \&
  {Avila-Reese}(2012){Rodr{\'{\i}}guez-Puebla}, {Drory}, \&
  {Avila-Reese}}]{RDA12}
{Rodr{\'{\i}}guez-Puebla} A., {Drory} N., {Avila-Reese} V., 2012, \apj, 756, 2

\bibitem[{{Sales} {et~al}\mbox{.}(2013){Sales}, {Wang}, {White}, \&
  {Navarro}}]{Sales+2013}
{Sales} L.~V., {Wang} W., {White} S.~D.~M., {Navarro} J.~F., 2013, \mnras, 428,
  573

\bibitem[{{Springel} {et~al}\mbox{.}(2008){Springel}, {Wang}, {Vogelsberger},
  {Ludlow}, {Jenkins}, {Helmi}, {Navarro}, {Frenk}, \& {White}}]{Springel+2008}
{Springel} V. {et~al.}, 2008, \mnras, 391, 1685

\bibitem[{{Stanimirovi{\'c}}, {Staveley-Smith} \&
  {Jones}(2004){Stanimirovi{\'c}}, {Staveley-Smith}, \&
  {Jones}}]{Stanimirovic+2004}
{Stanimirovi{\'c}} S., {Staveley-Smith} L., {Jones} P.~A., 2004, \apj, 604, 176

\bibitem[{{Strigari} {et~al}\mbox{.}(2007){Strigari}, {Bullock}, {Kaplinghat},
  {Diemand}, {Kuhlen}, \& {Madau}}]{Strigari+2007}
{Strigari} L.~E., {Bullock} J.~S., {Kaplinghat} M., {Diemand} J., {Kuhlen} M.,
  {Madau} P., 2007, \apj, 669, 676

\bibitem[{{Strigari}, {Frenk} \& {White}(2010){Strigari}, {Frenk}, \&
  {White}}]{Strigari+2010}
{Strigari} L.~E., {Frenk} C.~S., {White} S.~D.~M., 2010, \mnras, 408, 2364

\bibitem[{{Strigari} \& {Wechsler}(2012)}]{Strigari+2012}
{Strigari} L.~E., {Wechsler} R.~H., 2012, \apj, 749, 75

\bibitem[{{Tamm} {et~al}\mbox{.}(2012){Tamm}, {Tempel}, {Tenjes}, {Tihhonova},
  \& {Tuvikene}}]{Tamm+2012}
{Tamm} A., {Tempel} E., {Tenjes} P., {Tihhonova} O., {Tuvikene} T., 2012, \aap,
  546, A4

\bibitem[{{Tinker} {et~al}\mbox{.}(2008){Tinker}, {Kravtsov}, {Klypin},
  {Abazajian}, {Warren}, {Yepes}, {Gottl{\"o}ber}, \& {Holz}}]{Tinker+2008}
{Tinker} J., {Kravtsov} A.~V., {Klypin} A., {Abazajian} K., {Warren} M.,
  {Yepes} G., {Gottl{\"o}ber} S., {Holz} D.~E., 2008, \apj, 688, 709

\bibitem[{{Tollerud} {et~al}\mbox{.}(2011){Tollerud}, {Boylan-Kolchin},
  {Barton}, {Bullock}, \& {Trinh}}]{Tollerud+2011}
{Tollerud} E.~J., {Boylan-Kolchin} M., {Barton} E.~J., {Bullock} J.~S., {Trinh}
  C.~Q., 2011, \apj, 738, 102

\bibitem[{{Vogelsberger}, {Zavala} \& {Loeb}(2012){Vogelsberger}, {Zavala}, \&
  {Loeb}}]{Vogelsberger+2012}
{Vogelsberger} M., {Zavala} J., {Loeb} A., 2012, \mnras, 423, 3740

\bibitem[{{Wang} {et~al}\mbox{.}(2012){Wang}, {Frenk}, {Navarro}, {Gao}, \&
  {Sawala}}]{Wang+2012b}
{Wang} J., {Frenk} C.~S., {Navarro} J.~F., {Gao} L., {Sawala} T., 2012, \mnras,
  424, 2715

\bibitem[{{Wang} \& {White}(2012)}]{Wang+2012}
{Wang} W., {White} S.~D.~M., 2012, \mnras, 424, 2574

\bibitem[{{Wolf} \& {Bullock}(2012)}]{Wolf+2012}
{Wolf} J., {Bullock} J.~S., 2012, ArXiv e-prints

\bibitem[{{Yang}, {Mo} \& {van den Bosch}(2009){Yang}, {Mo}, \& {van den
  Bosch}}]{YMB09}
{Yang} X., {Mo} H.~J., {van den Bosch} F.~C., 2009, \apj, 695, 900

\bibitem[{{Yang} {et~al}\mbox{.}(2007){Yang}, {Mo}, {van den Bosch},
  {Pasquali}, {Li}, \& {Barden}}]{Yang+2007}
{Yang} X., {Mo} H.~J., {van den Bosch} F.~C., {Pasquali} A., {Li} C., {Barden}
  M., 2007, \apj, 671, 153

\bibitem[{{Yang} {et~al}\mbox{.}(2012){Yang}, {Mo}, {van den Bosch}, {Zhang},
  \& {Han}}]{Yang2012}
{Yang} X., {Mo} H.~J., {van den Bosch} F.~C., {Zhang} Y., {Han} J., 2012, \apj,
  752, 41

\bibitem[{{Yniguez} {et~al}\mbox{.}(2013){Yniguez}, {Garrison-Kimmel},
  {Boylan-Kolchin}, \& {Bullock}}]{Yniguez+2013}
{Yniguez} B., {Garrison-Kimmel} S., {Boylan-Kolchin} M., {Bullock} J.~S., 2013,
  ArXiv e-prints

\bibitem[{{Zavala}, {Vogelsberger} \& {Walker}(2013){Zavala}, {Vogelsberger},
  \& {Walker}}]{Zavala+2013}
{Zavala} J., {Vogelsberger} M., {Walker} M.~G., 2013, \mnras, 431, L20

\bibitem[{{Zolotov} {et~al}\mbox{.}(2012){Zolotov}, {Brooks}, {Willman},
  {Governato}, {Pontzen}, {Christensen}, {Dekel}, {Quinn}, {Shen}, \&
  {Wadsley}}]{Zolotov+2012}
{Zolotov} A. {et~al.}, 2012, \apj, 761, 71

\end{thebibliography}

\end{document}